\documentclass[prl,showpacs,showkeys,twocolumn]{revtex4}
\usepackage{bm}
\begin{document}
%----------------------------------------------------------------------------
\title{Conservation Laws in ``Doubly Special Relativity''}
%----------------------------------------------------------------------------
\author{Simon Judes}
\email{sj2057@columbia.edu} \affiliation{Columbia University, New
York, NY 10027, U.S.A.}
%--------------------------------------------------
\author{Matt Visser}
\email{matt.visser@vuw.ac.nz}
%\homepage{http://www.mcs.vuw.ac.nz/~visser}
\affiliation{School of Mathematical and Computing Sciences,
Victoria University of Wellington, New Zealand\\}
%-------------------------------------------------------------------------
\date{Revised 24 December 2002; \LaTeX-ed \today}
%-------------------------------------------------------------------------
\bigskip
%-------------------------------------------------------------------------
\begin{abstract}
%-------------------------------------------------------------------------
  Motivated by various theoretical arguments that the Planck energy
  ($E_\mathrm{Planck} \sim 10^{19}$ GeV) should herald departures from
  Lorentz invariance, and the possibility of testing these
  expectations in the not too distant future, two so-called ``Doubly
  Special Relativity'' theories have been suggested --- the first by
  Amelino-Camelia (DSR1) and the second by Smolin and Magueijo (DSR2).
  These theories contain two fundamental scales --- the speed of light
  and an energy usually taken to be $E_\mathrm{Planck}$. The symmetry
  group is still the Lorentz group, but in both cases acting
  nonlinearly on the energy-momentum sector. Since energy and momentum
  are no longer additive quantities, finding their values for
  composite systems (and hence finding appropriate conservation laws)
  is a nontrivial matter. Ultimately it is these possible deviations
  from simple linearly realized relativistic kinematics that provide
  the most promising observational signal for empirically testing
  these models. Various investigations have narrowed the conservation
  laws down to two possibilities per DSR theory. We derive unique
  exact results for the energy-momentum of composite systems in both
  DSR1 and DSR2, and indicate the general strategy for arbitrary
  nonlinear realizations of the Lorentz group.
%-------------------------------------------------------------------------
\end{abstract}
%-------------------------------------------------------------------------
\pacs{gr-qc/0205067}
%-------------------------------------------------------------------------
%-------------------------------------------------------------------------
\maketitle
%-------------------------------------------------------------------------
\input epsf
%----------------------------------------------------------------
% Local defines
%----------------------------------------------------------------
\def\bibnamefont{ }
\def\half{{1\over2}}
\def\L{{\mathcal L}}
\def\S{{\mathcal S}}
\def\d{{\mathrm{d}}}
\def\etal{{\emph{et al.}}}
\def\det{{\mathrm{det}}}
\def\tr{{\mathrm{tr}}}
\def\ie{{\emph{i.e.}}}
\def\eg{{\emph{e.g.}}}
\def\im{{\rm i}}
\def\bnabla{\mbox{\boldmath$\nabla$}}
\def\x{{\mathbf x}}
\def\aka{{\emph{aka}}}
\def\Choose#1#2{{#1 \choose #2}}
\def\etc{{\emph{etc.}}}
\def\Hospital{H\^opital}
\def\P{{\mathcal P}}
%----------------------------------------------------------------
\def\HRULE{{\bigskip\hrule\bigskip}}
%----------------------------------------------------------------
\def\be{\begin{equation}}
\def\ee{\end{equation}}
%-----------------------------------------------------------------
\def\bea{\begin{eqnarray}}
\def\eea{\end{eqnarray}}
%-----------------------------------------------------------------

%-------------------------------------
\noindent\underline{\noindent\emph{Background:}}
%-------------------------------------
Observations of very high energy cosmic rays, above the
expected ``GZK cutoff'' due to interaction with microwave
background radiation~\cite{GZK1,GZK2}, have precipitated a
surge of interest in possible violations of Lorentz invariance.
Encouragingly it appears that this phenomenon may furnish
experimental tests of some suggested theories of quantum
gravity~\cite{A-CEMNSarkar,ColemanGlashow,OBC,Kifune,Kluzniak,ACPiran}.
For a review, see~\cite{Sarkar}.  One type of Lorentz violating
theory is known as ``Doubly Special Relativity'' after
Amelino-Camelia~\cite{AC3}, who has suggested a specific example
of a DSR theory~\cite{DSR1}. Smolin and Magueijo have suggested
another~\cite{DSR2} in a paper in which they argued that any DSR
transformation group \emph{must} be a nonlinear realization of the
Lorentz group --- because that is the only suitable 6 parameter
extension of SO(3) --- the group of spatial rotations.
Unlike ordinary special relativity, in DSR the transformation
properties of energy and momentum need not be the same as those of the
space-time coordinates. Many investigations have been limited to the
energy-momentum sector~\cite{DSR1,AC3}. One approach that deals with
space-time as well (it it presently unclear if there are others) is in
terms of the $\kappa$--Poincar\'e algebra --- a deformation of the
Poincar\'e algebra~\cite{LukNowRueggTolstoy,Ruegg}. The algebras
obeyed by the DSR1 and DSR2 Lorentz generators are known to be just
such nonlinear deformations~\cite{DSR2,KG-Nowak,LukNow} of the
$\kappa$--Lorentz subalgebra --- DSR1 corresponding to the so-called
``bi-crossproduct basis''.  Because there is still some controversy
and uncertainty regarding the issue of whether or not all DSR theories
are \emph{necessarily} $\kappa$--Poincar\'e theories, we will stay in
momentum space and deal only with general features of arbitrary
nonlinear representations of the Lorentz group~\cite{Nature}.

To find conservation laws, two distinct approaches have been used.
One method~\cite{LukNow,KG-Nowak} is to investigate the nature of the
nonlinear realization of the symmetry group instantiated by the DSR
transformations and use its properties as constraints on the
conservation laws for composite systems. The
alternative~\cite{DSR1,AC3} is to work directly with the
transformation equations and to apply physically intuitive
restrictions to deduce the laws. Through a combination of these two
techniques, the number of possible conservation laws for DSR1 and DSR2
has been reduced to two. We continue along the lines of the second
method, and find that it is possible to uniquely identify the
conservation laws for \emph{any} DSR theory by applying seemingly
reasonable physical principles.
We give exact results for the total energy and momentum of a composite
system in both DSR1 and DSR2. Because these formulae implicitly
control particle production thresholds they are critically important
in assessing phenomenological attempts to place observational
constraints on the DSR
theories~\cite{Jacobson,Liberati,Sarkar,AC:IUCCA,Major}.

%-------------------------------------
\noindent\underline{\noindent\emph{General rules:}}
%-------------------------------------
Since a DSR symmetry group is simply a nonlinear realization of
the Lorentz group~\cite{KG-Nowak,LukNow,DSR2}, we can find
functions of the physical energy-momentum $P_4=(E,p)$ which
transform like a Lorentz 4--vector. These we will call the
pseudo-energy-momentum $\P_4=(\epsilon,\pi)$, but it should not be
thought that these necessarily have immediate physical
significance. We have:
\be
P_4 = F(\P_4); \qquad\qquad \P_4 = F^{-1}(P_4).
\ee
The function $F$ and its inverse $F^{-1}$ are in general
complicated nonlinear functions from $\Re^4$ to $\Re^4$, but both
of course reduce to the identity in the limit where energies and
momenta are small compared to the DSR scale. The Lorentz
transformations act on the auxiliary variables in the normal
linear manner:
\( (\epsilon';\pi') = \L\; (\epsilon;\pi). \)
where $\L$ is the usual Lorentz transformation, boosting from the
unprimed coordinates to the primed coordinates. The boost operator for
the physical energy and momentum $(E,p)$ we call $L$, and is given by
the composition:
\be P_4' = L(P_4) = [\,F\circ \L\circ F^{-1}\,](P_4). \ee Now $\L$
and $F$ uniquely determine the nonlinear Lorentz transformation
$L$; however $\L$ and $L$ [more precisely, $L(\L)$] do not
uniquely determine the function $F$ --- there is an overall
multiplicative ambiguity which must be dealt with using the
dispersion relation:
\be [\epsilon(E,p)]^2 - [\pi(E,p)]^2 = \mu_0^2. \ee
Here $\mu_0$ is simply the Lorentz invariant constructed from
$\epsilon$ and $\pi$ (the Casimir invariant); not to be confused with
the rest energy. In terms of the rest energy $m_0$, obtained by going
to a Lorentz frame in which the particle is at rest,
\( \mu_0 = \epsilon(m_0,0). \)
The combination of $L(\L)$ and $\mu_0(m_0)$ is now sufficient to
pin down $F$ completely.

In the linear representation, kinematic quantities such as total
energy can be defined in the usual fashion
\be
\P_4^{tot} = \sum_i \P_4^i.
\label{E:Ptot}
\ee
Calculating the total physical 4-momentum is then straightforward:
\be \label{E:generic} P_4^{tot} = F\left( \sum_i F^{-1}(P_4^i)
\right).
\label{E:ptot}
\ee
This is the quantity that will be conserved in collisions.
Calculating it is simply a matter of finding $F$ and its inverse.

%-------------------------------------
\noindent\underline{\noindent\emph{Variant conservation laws:}}
%-------------------------------------
The choice in equation (\ref{E:Ptot}), and so implicitly in equation
(\ref{E:ptot}), can be uniquely characterized by saying that the
general composition of 4-momenta is based on iterating an associative
symmetric binary function.

If the general composition law were not based on iterating a binary
function, then one would need to postulate an infinite tower of
distinct composition laws for $2, 3, 4, \dots, n, \dots$ particles.
Such a situation would create serious difficulties in the
interpretation of quantum field theories: For instance,
energy-momentum conservation at each vertex of a Feynman diagram would
now depend in an essentially arbitrary way on a particular time-slice
through the diagram and the energy-momenta of all other particles in
the diagram as they cross that time-slice. Perhaps worse, every time a
dressed particle were to either emit or absorb a virtual particle one
would have to completely recalculate the energy-momentum for the entire
virtual cloud.

If the binary function were not symmetric, one could
(simply by changing the order in which one chooses to list the
particles) construct symmetric and anti-symmetric combinations,
leading to two separate conservation laws that would over-constrain
the collision (unless, of course, the anti-symmetric law happens to be
trivial --- but that implies a symmetric binary function).

Finally, if the binary function were not associative, then the
energy-momentum of a composite system would depend not only on the
constituents of the system, but also on the manner in which the system
is aggregated out of subsystems --- an option that is at best extremely
unnatural.

The initial investigations into energy and momentum of composite
systems in DSR~\cite{DSR1} proceeded only on the requirement that the
law of energy-momentum conservation had to be covariant with respect
to the DSR transformation.  The insufficiency of this requirement is
manifest when we consider that the following definition:
\be
\P_4^{tot} = \sum_i \nu_i \; \P_4^i.
\ee
produces a covariant conservation law for \emph{arbitrary} $\nu_i$.
Symmetry, which is required to prevent over-determining the
energy-momentum in a collision, implies that:
\be
\P_4^{tot} = \nu \; \sum_i \P_4^i.
\ee
If this is to arise from iterating a two-particle composition law we
need $\P_4^{\{12\}} = \nu (\P_4^1+\P_4^2)$.  But now for a
three-particle system, associativity implies
\be
%\P_4^{123} =
\nu [\nu (\P_4^{\rm 1} +\P_4^{\rm 2}) + \P_4^{\rm 3}]
=
\nu [\P_4^{\rm 1} + \nu (\P_4^{\rm 2} +\P_4^{\rm 3})]
\ee
Therefore $\nu = \nu^2$, implying either $\nu=1$ or $\nu=0$.  This
argument gives the same result as that used by Lukierski and
Nowicki~\cite{LukNow} to reduce the number of possible laws to two. In
fact, their ``symmetric'' and ``non-symmetric'' laws are just the
$\nu=1$ and $\nu=0$ cases respectively.  The $\nu=1$ solution is
clearly unproblematic. However, what is not evident from the group
theoretic analysis of~\cite{LukNow}, and is evident from the current
approach, is the rather odd nature of the case where $\nu=0$.  Taken
straightforwardly, it must be false, implying that for any number of
particles
\be
\P_4^{tot} = \vec 0; \qquad P_4^{tot} = F(\vec 0).
\ee
Thus $\nu=0$ is clearly unphysical and we are forced to adopt the
intuitive choice of $\nu=1$.

We feel that more drastic possibilities~\cite{Nature}, based on
abandoning notions of an iterated associative symmetric binary
composition law are strongly disfavoured, and will not pursue such
options in this Letter.

%-------------------------------------
\noindent\underline{\noindent\emph{DSR2:}}
%-------------------------------------
This model~\cite{DSR2} is completely characterized by the equation
\be \P_4 \equiv (\epsilon;\pi) = F^{-1}(P_4) =
{(E;p)\over1-\lambda E}. \ee
(In model building one typically takes $\lambda =
1/E_{\mathrm{Planck}}$; but we will leave $\lambda$ as an
arbitrary parameter with dimensions $[E]^{-1}$.)  The inverse
mapping is easily established to be
\be P_4 \equiv (E; p) = F(\P_4) = {(\epsilon;\pi)\over1+\lambda
\epsilon}. \ee
The total physical 4-momentum is easily calculated. First observe
that for the pseudo-momenta
\be
\label{E:dsr2-eps}
\epsilon_{tot} = \sum_i {E_i\over1-\lambda
E_i};
\qquad \label{E:dsr2-pi}
\pi_{tot} =  \sum_i {p_i\over1-\lambda E_i}.
\ee
Then
\be
\label{E:dsr2-E}
E_{tot} = {\sum_i {E_i/(1-\lambda E_i)} \over
1 + \lambda \sum_i {E_i/(1-\lambda E_i)} } .
\ee
and
\be
\label{E:dsr2-p} p_{tot} = {\sum_i {p_i/(1-\lambda E_i)} \over
1 + \lambda \sum_i {E_i/(1-\lambda E_i)} }.
\ee

Within the framework of DSR2 this result is \emph{exact} for all
$\lambda$. To first-order in $\lambda$:
\be E_{tot} = \sum_i E_i - \lambda \sum_{i\neq j} E_i E_j +
O(\lambda^2), \ee
\be p_{tot} = \sum_i p_i - \lambda \sum_{i\neq j} p_i E_j +
O(\lambda^2). \ee
For the case of two particles, the above formulae reduce to the
so-called `mixing laws' --- one of the possibilities mentioned by
Amelino-Camelia \emph{et al.}~\cite{AC3}.

We also mention in passing that the exact dispersion relation for
DSR2 is
\be {E^2-p^2\over(1-\lambda E)^2} = \mu_0^2 =
{m_0^2\over(1-\lambda m_0)^2}. \ee
This can be rearranged as
\be p^2 = E^2 - m_0^2 \left( {1-\lambda E\over 1-\lambda m_0}
\right)^2. \ee
Solving the quadratic for $E$, and choosing the physical root
\be E = { \sqrt{(1-2\lambda m_0)[m_0^2+(1-\lambda m_0)^2
p^2]+\lambda^2 m_0^4} - \lambda m_0^2 \over 1-2\lambda m_0 }. \ee

%-------------------------------------
\noindent\underline{\noindent\emph{DSR1:}}
%-------------------------------------
For DSR1 the basic principles are the same but the algebra is somewhat
messier.  It is convenient to consider a particle at rest, with rest
energy $m_0$, and then boost using a rapidity parameter $\xi$. The
defining relationships for DSR1 can then be put in the
form~\cite{DSR1}
\be
e^{\lambda E} =
e^{\lambda m_0}
\left(1+\sinh(\lambda m_0) \; e^{-\lambda m_0} \; [\cosh\xi-1] \right),
\ee
and
\be p = {1\over\lambda} \;
{ \sinh(\lambda m_0) \; e^{-\lambda m_0} \; \sinh\xi
\over
1+\sinh(\lambda m_0) \; e^{-\lambda m_0} \; [\cosh\xi-1] }.
\ee
(These expressions are equivalent to knowing the nonlinear Lorentz
transformations $L$ as a function of rapidity $\xi$.)  This can
easily be inverted to give expressions for the rapidity
\be \cosh \xi = {e^{\lambda E} - \cosh(\lambda m_0)\over
\sinh(\lambda m_0)}; \quad \sinh \xi = {\lambda \; p \; e^{\lambda
E}\over \sinh(\lambda m_0)}. \ee Making use of the identity
$\cosh^2\xi-\sinh^2\xi=1$ gives the DSR1 dispersion relation in
the particularly nice form
\be
\cosh(\lambda E) = \cosh(\lambda m_0) + {1\over2} \lambda^2
p^2 e^{\lambda E}.
\ee
Comparison with the standard form of the dispersion relation now
fixes the rest energy in terms of the Casimir invariant
\be
\cosh(\lambda m_0) = 1 + {1\over2} \lambda^2 \mu_0^2;
\qquad
\mu_0 = {2\sinh(\lambda m_0/2)\over\lambda}. \ee
This now fixes the linear representation completely. In terms of
the physical energy-momenta
\be \epsilon = \mu_0  \cosh\xi = {e^{\lambda E} - \cosh(\lambda
m_0)\over \lambda\cosh(\lambda m_0/2)} , \ee
and
\be \pi = \mu_0 \sinh\xi = {p \; e^{\lambda E}\over\cosh(\lambda
m_0/2)} . \ee Conversely, the inverse mappings yielding physical
energy-momenta in terms of auxiliary energy-momenta are
\bea
E &=& {1\over\lambda} \ln\left[{\lambda\epsilon\cosh(\lambda
m_0/2)}+\cosh(\lambda m_0) \right],
\nonumber\\
&=& {1\over\lambda} \ln\left[1+\lambda\epsilon\sqrt{1+{\lambda^2
\mu_0^2\over4}} +{\lambda^2 \mu_0^2\over2} \right],
\nonumber\\
&=& {1\over\lambda} \ln\left[1+\lambda\epsilon\sqrt{1+{\lambda^2
(\epsilon^2-\pi^2)\over4}} +{\lambda^2 (\epsilon^2-\pi^2)\over2}
\right],
\nonumber\\
&&
\eea
and
\bea
p &=& \pi \cosh(\lambda m_0/2) e^{-\lambda E},
\nonumber\\
&=& { \pi \sqrt{1+{\lambda^2 \mu_0^2\over4}} \over
1+\lambda\epsilon\sqrt{1+{\lambda^2 \mu_0^2\over4}}+{\lambda^2
\mu_0^2\over2} },
\nonumber\\
&=& { \pi \sqrt{1+{\lambda^2 (\epsilon^2-\pi^2)\over4}} \over
1+\lambda\epsilon\sqrt{1+{\lambda^2  (\epsilon^2-\pi^2)\over4}}
+{\lambda^2 (\epsilon^2-\pi^2) \over2} }.
\eea

To calculate total energy and momentum of a collection of
particles we now first calculate auxiliary quantities
\be \label{E:dsr1-eps} \epsilon_{tot} = \sum_i {e^{\lambda E_i} -
\cosh(\lambda m_{0,i})\over \lambda\cosh(\lambda m_{0,i}/2)}, \ee
\be \label{E:dsr1-pi} \pi_{tot} = \sum_i {p_i \; e^{\lambda
E_i}\over\cosh(\lambda m_{0,i}/2)}, \ee
and then use these to calculate the physical quantities \bea
\label{E:dsr1-E} E_{tot} &=& {1\over\lambda}
\ln\bigg[1+\lambda\epsilon_{tot}\sqrt{1+{\lambda^2
(\epsilon_{tot}^2-\pi_{tot}^2)\over4}}
\nonumber\\
&& \qquad\qquad +{\lambda^2 (\epsilon_{tot}^2-\pi_{tot}^2)\over2}
\bigg], \eea
\bea \label{E:dsr1-p} p_{tot} &=& { \pi_{tot} \sqrt{4+{\lambda^2
(\epsilon_{tot}^2-\pi_{tot}^2)}} \over
2+\lambda\epsilon_{tot}\sqrt{4+{\lambda^2
(\epsilon_{tot}^2-\pi_{tot}^2)}} +{\lambda^2
(\epsilon_{tot}^2-\pi_{tot}^2)} }.
\nonumber\\
&& \eea These formulae provide explicit (albeit complicated)
expressions for the total physical energy and momenta in the DSR1
model in terms of the individual physical energy, momenta, and
rest energies; note that the formulae are \emph{exact} for
arbitrary $\lambda$.

To first order:
\be E_{tot} = \sum_i E_i - {1\over2}\lambda \sum_{i\neq j} p_i p_j
+ O(\lambda^2), \ee
\be p_{tot} = \sum_i p_i - \lambda \sum_{i\neq j} p_i E_j
+O(\lambda^2). \ee
For two particles, these too reduce to equations already in the
literature~\cite{DSR1,LukNow}.

%-------------------------------------------------------
\noindent\underline{\noindent\emph{Discussion:}}
%------------------------------------------------------
The key result of this note is the identification of appropriate laws
of conservation of energy and momentum in generic DSR theories,
embodied in the general formula (\ref{E:generic}), together with the
specific applications to DSR2 in equations
(\ref{E:dsr2-E})---(\ref{E:dsr2-p}), and to DSR1 in equations
(\ref{E:dsr1-eps})---(\ref{E:dsr1-p}).  Ultimately the general formula
(\ref{E:generic}) is more important: There are many ways of distorting
the Lorentz group, and this formula applies to all of them --- this
makes it clear that the distortion of dispersion relations, the
existence of unexpected thresholds, and the somewhat unexpected
subtleties hiding in the conservation laws are generic to all
nonlinear realizations of the Lorentz group, no matter how they are
obtained. It is these possible deviations from simple linearly
realized relativistic kinematics that provide the most promising
observational signal for empirically testing these
models~\cite{Jacobson,Liberati,Sarkar}.

%-------------------------------------
\noindent\underline{\noindent\emph{Acknowledgements:}}
%-------------------------------------
One of us (S. Judes) wishes to thank Subir Sarkar for advice and
support.
%----------------------------------------------------------------

%----------------------------------------------------------------
%----------------------------------------------------------------
\end{document}